\documentclass[manuscript]{acmart}

\AtBeginDocument{}

\setcopyright{acmlicensed}
\copyrightyear{2024}
\acmYear{2024}
\acmDOI{XXXXXXX.XXXXXXX}

\acmConference[Conference Name]{Conference Name}{Month, Year}{Location}
\acmBooktitle{Proceedings of the Conference Name (Conference Name '24), Month, Year, Location}
\acmISBN{978-1-4503-XXXX-X/24/08}

\usepackage{amssymb}
\usepackage{amsmath}
\usepackage{algorithm}
\usepackage{algorithmic}
\usepackage{graphicx}
\usepackage{textcomp}
\usepackage{xspace}
\usepackage{listings}
\usepackage{beramono}
\usepackage{tikz}
\usetikzlibrary{positioning}
\usepackage{pgfplots}
\pgfplotsset{compat=1.18} \usepackage{ifthen}
\usepackage{wrapfig}
\usepackage{hyperref}
\usepackage{cleveref}

\usepackage{enumitem}

\newcommand*\LSTfont{\small\ttfamily}
\newcommand{\todo}[1]{{\color{red}\bfseries [[#1]]}}

\ifdefined\notodocomments
  \renewcommand{\todo}[1]{\relax}
\fi

\newcommand{\NullGTN}{NullGTN\xspace}
\newcommand{\nullgtn}{\NullGTN\xspace}

\newcommand{\napast}{NaP-AST\xspace}
\newcommand{\napasts}{NaP-ASTs\xspace}

\newif\ifanonymous
\anonymoustrue

\newcommand{\anonurl}[1]{\ifanonymous URL removed for anonymity.\else\url{#1}\fi}

\def\|#1|{\mathid{#1}}
\newcommand{\mathid}[1]{\ensuremath{\mathit{#1}}}
\def\<#1>{\lstinline{#1}}

\newlist{researchquestions}{enumerate}{1}
\setlist[researchquestions]{label*=\textbf{RQ\arabic*}}

\begin{document}

\title{Nullability Type Inference via Machine Learning}

\author{Kazi Amanul Islam Siddiqui}
\affiliation{\institution{New Jersey Institute of Technology}
  \city{Newark}
  \state{NJ}
  \country{USA}}
\email{ks225@njit.edu}

\author{Martin Kellogg}
\affiliation{\institution{New Jersey Institute of Technology}
  \city{Newark}
  \state{NJ}
  \country{USA}}
\email{martin.kellogg@njit.edu}

\begin{abstract}
Pluggable type systems improve code safety by adding type qualifiers, but deploying them in legacy Java projects is difficult because annotations must be written by hand. This paper investigates using machine learning to infer type qualifiers. We evaluate graph models—Graph Convolutional Networks (GCNs) and Graph Transformer Networks (GTNs)—and large language models (LLMs). For graphs, we test Type Dependency Graph (TDG) encoding from prior work and introduce a novel encoding, Name-augmented, Pruned Abstract Syntax Tree (NaP-AST), embedding minimal dataflow hints. Models are validated on 9 open-source projects human annotated for the NullAway nullness checker. The GTN with NaP-AST achieves the best downstream impact, cutting checker warnings by 69\%, with 69\% recall of human-written annotations and 39\% precision. The strongest LLM shows the opposite trade-off: 91\% precision but only 21\% recall, still reducing warnings by 61\%. Regarding generalization, we conducted an experiment to estimate the number of classes required in the training set to achieve good performance: performance improves up to roughly 16K annotated classes and then degrades from overfitting around 22K. \end{abstract}




\maketitle

\section{Introduction}
\label{sec:intro}
When programmers write code, they are often aware of semantic properties that a variable
must adhere to but which are not checked by the type system of their programming language---for example,
that a variable should only hold non-null values.
Pluggable type systems~\cite{FosterFFA99,Bracha2004} add custom \emph{type qualifiers} to a language.
A pluggable typechecker statically checks these user-defined qualifiers
and warns about potential violations at compile time. For example, PUnits~\cite{XiangLD2020}
adds qualifiers for scientific units, while Uber's NullAway~\cite{BanerjeeCS2019}, Meta's NullSafe~\cite{PianykhZL2022},
and the Checker Framework's Nullness Checker~\cite{PapiACPE2008,DietlDEMS2011}
enforce programs do not crash due to null-pointer dereferences with an \<\@Nullable> qualifier.
A key barrier to bringing legacy codebases into compliance with a pluggable typechecker is that developers must write
these qualifiers into their code to see the full benefit.
The state of the practice when deploying a new typechecker on a legacy codebase
remains manual annotation: that is, type qualifiers are written by
hand. The high cost of this manual process discourages developers from using these
systems, despite their promise to improve code quality. This paper addresses the problem
of bringing a legacy codebase into compliance with a null-safety typechecker by automatically inferring type qualifiers.
While our primary focus is on nullability for data availability reasons,
we propose a framework that is theoretically adaptable to other finite pluggable type systems,
if sufficient training data were available.

Recently, researchers have proposed using machine learning to solve a related, but different, problem:
inferring type annotations for Python or JavaScript programs~\cite{peng2023statistical,wei2023typet5,peng2023generative,PengGLGLZL2022,mir2022type4py,cui2021pyinfer}.
This problem shares a key feature with inferring type qualifiers for a legacy codebase:
in both cases, the programs were written \emph{without feedback from the typechecker}, so
we expect them not to typecheck cleanly (with any set of annotations).
Experimental results suggest that these techniques outperform traditional deductive type inference techniques.
Moreover, large language models (LLMs) have recently shown promise on a variety of coding tasks~\cite{abhik-agent-opinion,miserendino2025swe,mundler2024swt,bouzenia2025repairagent,yang2024swe,ozkaya2023application, wu2023large, xiong2023program, ross2023programmer}.
This paper's goal is to investigate if similar approaches can work for inferring pluggable types.

However, techniques for inferring types in Python or JavaScript programs
are not immediately transferable to our setting, because of three key differences.
First, when inferring types for Python or JavaScript, the type system is fixed
(it enforces ``traditional'' type safety: objects do not try to perform invalid operations).
A key advantage of pluggable type systems is their customizability:
each type system enforces its own rules. So, we desire a way to customize our inference approach
that does not require us to start over from scratch for each new type system.
Second, when inferring types for e.g., Python, the space of possible types is infinite
and has a long tail, because of the presence of user-defined types.
Prior works focus on solving this so-called \emph{rare type problem}; for example,
by encoding programs using a \emph{Type Dependency Graph} (\emph{TDG})~\cite{peng2022static} data structure
which is specialized to help the model find rare types.
However, while the tail of types is long, the set of \emph{locations} that need a type annotation
is not: in prior work, the locations where a type should be written are fixed in advance.
For most pluggable type systems, these problems are reversed: the hardest problem
is not choosing which type to write, but at which places to write a type. For example,
for a nullability type system, an annotator usually only needs to write either \<\@Nullable>
or \<\@NonNull> (and the pluggable type system interprets the \emph{lack} of a qualifier as
the other), so the problem that an inference tool needs to solve is \emph{where} to write
a qualifier rather than \emph{which} type to write. This difference makes the encodings
from prior works (e.g., the TDG) ill-suited to our setting.
Third, machine learning is data hungry: approaches for popular languages like Python can assume
the existence of a large corpus of human-written types in open-source software. The customizability of pluggable types is a double-edged sword: many pluggable type systems
are highly-specialized and therefore have relatively few users. So, to apply machine learning
techniques to the pluggable type inference setting, we need to understand how much data is actually
required to make them effective.
These differences inspired us to investigate two high-level research questions:
\begin{researchquestions}
\item What kind of machine-learning model architecture is suitable for inferring pluggable types, and how does that model differ from
  those in prior type inference works for other languages?
\item How much data is needed to make that model effective at inferring qualifiers in practice?
\end{researchquestions}
The results of our investigation of these questions allows us to propose a general pipeline
for training an ML model to infer type qualifiers for a given pluggable type system.

State-of-the-art models~\cite{peng2023statistical,wei2023typet5} for inferring types in Python/JavaScript use
diverse model architectures, including graph and text-based models:
Typilus~\cite{allamanis2020typilus} learns a type space,
STIR~\cite{peng2023statistical} uses a BiLSTM for incomplete code,
and TypeT5~\cite{wei2023typet5} uses a pre-trained seq2seq model.
The differences with the pluggable types setting, however, inspired us to investigate
\textbf{RQ1} by evaluating a set of models used in previous work.
We eschew time-series models like BiLSTM (which are effective for sequential
data, like recognizing the content of a video, but very expensive during training
and inference) because predicting where to write a type qualifier is more similar to a classification task.
We considered four models: two graph models and two text-based models.
For the graph models, we considered two encodings: the TDG encoding~\cite{peng2022static} from prior work
and a novel encoding that we developed specifically for pluggable type inference.
Our novel graph encoding methodology uses ablation studies to
retain information useful for predicting the type qualifiers of
a particular type system,
which results in a novel program encoding that we call a \emph{Name-augmented, Pruned Abstract Syntax Tree} (\napast), which we describe further in \cref{sec_meth}.
A \napast encodes the parts of the program's structure that are relevant to the type system of interest, along with
name information that ties related program elements together, into a graph that is a suitable
input to a graph-based model.
The first graph model is a baseline~\cite{wu2019simplifying}
graph convolutional network (GCN) model~\cite{zhang2019graph}.
The second is a more-capable graph transformer network (GTN) model~\cite{yun2019graph}, which extends
the GCN to reason about multiple edge types (i.e., hypergraphs).
Since the TDG encoding is not a heterogenous graph, we only use it with the GCN model; the \napast encoding
is a heterogenous graph, so it can benefit from the GTN model.
Both text-based models are large language models (LLMs), which we include because LLMs have performed
well on coding tasks recently. The first is CodeQwen1.5-7B-Chat~\footnote{https://huggingface.co/Qwen/CodeQwen1.5-7B-Chat}, an open-source model specialized
for coding tasks.
The second text-based model is GPT-4~\cite{achiam2023gpt}, a state-of-the-art\footnote{GPT-4 was OpenAI's flagship model when these experiments were conducted, though at the time of submission OpenAI recently (August 7, 2025) released its new GPT-5 model.} commercial LLM.

To evaluate these models and encodings, we conducted a feasibility study:
we trained models to predict where to place type qualifiers for one popular pluggable type system.
We would have preferred to repeat this experiment for more than one type system, to show the
generality of our approach, but a lack of publicly-available data prevents us from doing so;
we show this empiricially in \cref{sec:rq3-nonnull}.
Of the commonly used type qualifiers, those for nullability in Java are the most common.
We therefore conducted our feasibility study on nullness: if there is not enough data for Java nullness
type qualifiers to train an effective model, there is not enough data to train an effective model
for \emph{any} pluggable type system.
First, we collected a dataset from GitHub of Java classes, each of which uses one of
31 different \<@Nullable> annotations.
After cleaning, this dataset contains 32,370 classes.
We then used our methodology for building a \napast encoding to derive a \napast encoding for nullness.
Then, we trained the four models using this dataset:
the two graph models (a GCN and a GTN) using the derived \napast encoding, the GCN using the baseline TDG encoding
(which cannot benefit from the GTN, because it is a homogenous graph),
and the text-based LLMs (CodeQwen1.5-7B-Chat and GPT-4) using syntax-based code chunking.
We then compared their performance on 9 open-source projects that had previously been annotated for
the popular NullAway typechecker~\cite{BanerjeeCS2019}.
The GTN with the \napast encoding outperformed the other models we evaluated:
both GCN models have lower recall than \nullgtn; TDG GCN has higher precision than \nullgtn but fails to reduce warnings, while NullGCN trails on both precision and recall. LLMs can hallucinate edits at scale; CodeQwen attains very high precision (0.91) but low recall (0.21), and GPT-4 has high precision (0.76) with moderate recall (0.25).
Our best model (which we call \nullgtn) achieves
69\% recall of the human-written type qualifiers and 39\% precision;
its inferred annotations
eliminate 69\% of the NullAway warnings that a human would need to
triage (a proxy for human effort).

Our success for nullability suggests that our approach may be effective with other type qualifiers,
but unfortunately we lack enough data for any other type qualifier to test this hypothesis.
To assist future researchers or practitioners who want to train a similar model for
another type system (and to answer our \textbf{RQ2}),
we explore how much training data one needs for adequate performance.
We rebuilt our best model with random samples of the dataset and re-ran
the evaluation to show how much data it needs before it achieves good performance.
Around 16k human-annotated classes were needed: a large but not impossible-to-reach
number.
Our primary contributions are:
\begin{itemize}
\item A novel methodology to construct a graph encoding of programs for null safety (\napasts) 
  to train deep-learning graph models like GCNs and GTNs that can theoretically be extended to other finite type systems (\cref{sec_meth});
\item A dataset of 32,370 Java classes annotated with \<@Nullable> that enables our feasibility
  study (\cref{ssec_dp});
\item A comparison of three model architectures (GCN, GTN, and LLM) and two graph encodings (\napast and TDG) that shows that the GTN with the \napast encoding
  is best at inferring pluggable types in our feasibility study (\cref{sec:nullaway-eval}), with open-source models and data;
\item A study of the amount of data required by our best GTN model on the task of inferring type qualifiers for adequate performance (\cref{ssec:rq3}).
\end{itemize}
 \section{Background: Model Architectures}\label{sec:model}

This section describes the three considered model architectures: graph convolutional networks (GCNs),
graph transformer networks (GTNs), and large language models (LLMs).
We focus on their salient features with respect to type qualifier inference.

\subsection{Graph Models}

\subsubsection{GCN}
Graph Convolutional Networks (GCNs) are a class of neural networks for graph data.
Type qualifier prediction can be modeled as a graph data task
since ASTs can be represented as a graph.
GCNs use \emph{graph convolution}, a propagation rule that updates the feature representation of each node by aggregating features from its neighbors and the node itself.
For AST prediction, GCNs can be particularly effective, since they explicitly model local context:
ASTs are inherently hierarchical and structured, where the context around a node (e.g., its parents or children) significantly influences it.
GCNs leverage this local structure efficiently.
They are also efficient in learning spatial features, which have been helpful for ASTs~\cite{zhang2019novel}.\looseness=-1

A more specialized graph neural network model might outperform a GCN by
adapting to heterogeneous relationships.
If different types of connections between AST nodes (e.g., data flow vs. control flow) play a significant role, models within the broader GNN category designed to handle such heterogeneity might be more effective.
We investigate this possibility by comparing GCN with such a model, GTN.
Certain GNN architectures might also be better suited for incorporating global context or long-range dependencies within the AST, which could encode dataflow information.

\subsubsection{GTN}\label{sec:fastgtn}
GCNs assume their input graphs are homogeneous, i.e., they consist of one type of nodes and edges.
Using a model that supports heterogenous graphs lets us layer hints over the AST, which we use to add a layer connecting names across an AST to provide dataflow hints.

We use the FastGTN model~\cite{b11}, which uses the meta-path, a multi-hop sequence of connected nodes and edges in the graph with heterogeneous edge types. Instead of ignoring the edge type information, breaking down a heterogeneous graph into homogeneous graph by its meta-paths allows the model to preserve this information.
A Graph Transformer Network (GTN)~\cite{b11} learns the most informative meta-paths automatically for each task in an end-to-end manner by selecting adjacency matrices through a 1x1 convolution with weights from a softmax function. The GTN can learn arbitrary meta-paths with respect to edge types and path length.
GTNs consider multiple types of meta-paths simultaneously by generating multiple graph structures: they set the output channels of a 1x1 filter to the number of channels. After stacking $k$ GT layers, multi-layer GNNs are applied to each channel of the output tensor. The final node representations can be used for downstream tasks, by applying dense layers followed by a softmax layer to the node representations.

\begin{wrapfigure}{r}{0.5\textwidth}
  \centering
  \small
    \begin{tikzpicture}
\node[draw, rectangle, fill=cyan!30, minimum width=2cm, minimum height=1cm] (box1) {MethodDecl};
\node[draw, rectangle, fill=cyan!30, minimum width=2cm, minimum height=1cm, below right=of box1.south west, xshift=0cm, yshift=0.1cm] (box2) {SimpleName};
\node[draw, rectangle, fill=cyan!30, minimum width=2cm, minimum height=1cm, above right=of box1.north east, xshift=-1cm, yshift=0.1cm] (box5) {"f"};
\node[draw, rectangle, fill=cyan!30, minimum width=2cm, minimum height=1cm, below right=of box1.south east, xshift=1cm, yshift=0.1cm] (box3) {SimpleName};
\node[draw, rectangle, fill=cyan!30, minimum width=2cm, minimum height=1cm, right=of box1, xshift=1.5cm, yshift=0.1cm] (box4) {MethodCallExpr};

\draw[->] (box1) -- node[midway, left] {AST Child} (box2);
\draw[->] (box2) -- node[near end, left] {Name Parent} (box5);
\draw[->] (box5) -- node[near start, right] {Name Child} (box3);
\draw[->] (box3) -- node[midway, right] {AST Parent} (box4);
\end{tikzpicture}
    \caption{In heterogeneous graphs (i.e. having multiple edge types), a meta-path is a sequence of connected nodes and edges, where each edge has a distinct type.}
    \label{fig:metapaths}
\end{wrapfigure}
 
A weakness of GTNs is that they are not scalable because of the high computational cost and memory requirements when calculating and storing new adjacency matrices at each layer. To address this, FastGTNs implicitly transform the graph structures without storing the new adjacency matrices of meta-paths, avoiding the need for explicit multiplications of large adjacency matrices. This reduces the computational cost and memory usage while keeping FastGTNs mathematically identical to GTNs.
Even FastGTNs have serious scalability limitations, which motivates
our pruning strategies discussed in \cref{sec_meth}; note that the primary scalability
challenge for GTNs is fitting the $n$-by-$n$ adjacency matrix they require into memory
(where $n$ is the size of the input)---even with ``large'' inputs, both training
and prediction with FastGTNs is relatively quick (see further discussion in \cref{ssec:gtn}).

\subsubsection{GCN vs GTN}
The GTN is a generalization of a GCN that supports distinct edge types. This allows the GTN to distinguish the AST and the name layer.
While GCNs do not automatically support this heterogeneous operation,
they are a good baseline for studies exploring advanced graph learning techniques~\cite{wu2019simplifying}
because they can integrate local neighborhood information through the graph convolution operation.
\looseness=-1

\subsubsection{Models from Prior Work}
Although we did test the TDG encoding (\cref{sec:tdg}) from prior work,
we did not evaluate the Python-specific models with which it was originally
deployed, because they are not transferable to our setting.
In particular, the Typilus paper~\cite{allamanis2020typilus} used a custom ``TypeSpace'', metric-based meta-learning, and a Graph Neural Network (GNN) model
optimized for predicting rare types. None of this machinery is sensible or necessary
for predicting pluggable type qualifiers, which are typically drawn from a fixed
set---for example, in our feasibility study, from the set
containing \<@Nullable> and \<@NonNull>.

\subsection{Text-based Models}

\subsubsection{CodeQwen1.5-7B-Chat}
CodeQwen 1.5-7B-Chat\footnote{https://qwenlm.github.io/blog/codeqwen1.5/} is a fully open-source transformer (7 B params, 64 K context) trained on 3 T code tokens spanning 92 programming languages, and attains 83.5\% on HumanEval-plus and 78.7\% on MBPP-plus, outscoring other 7B code models~\footnote{https://huggingface.co/Qwen/CodeQwen1.5-7B-Chat}.
The model is fine-tuned with a synthetic instruction corpus that couples long-context code snippets with conversational tasks~\cite{bai2023qwen}.
Because CodeQwen 1.5-7B-Chat outperforms StarCoder 2, CodeLlama 7 B and DeepSeek-Coder on major code benchmarks, it is an appropriate open-source baseline.
The LLM approach leverages a vast knowledge base, ensuring a diverse understanding of coding patterns and practices, significantly beneficial for inferring type qualifiers by understanding the broader context and nuances in code. LLMs have recently been used to great effect in a number of software engineering domains~\cite{abhik-agent-opinion,miserendino2025swe,mundler2024swt,bouzenia2025repairagent,yang2024swe,ozkaya2023application, wu2023large, xiong2023program, ross2023programmer}.
Their limitation is that they may struggle with deep semantic understanding of code, as they are only predicting the next token in a sequence. Specifically, LLMs hallucinate about code structure.
These hallucinations add up when scaled to hundreds of thousands of lines of code.
While CodeQwen benefits from the 64 K-token window, it can still hallucinate nonsensical changes, like adding or removing import statements in large files. All current LLM systems hallucinate since their function is to predict the next token. To get around this limitation, we only asked the LLM ``yes or no'' questions, restricting its ability to change code (see \cref{sec:finetune} for more details).
In practice, there are other limits on LLMs, like token limits and cost.

\subsubsection{GPT-4o-mini}
GPT-4o-mini is an LLM that demonstrates exceptional performance in coding and natural language processing tasks, outperforming its predecessors and many contemporary models in various domains. For example, GPT-4o-mini achieves superior accuracy in programming competitions and exhibits strong cross-language translation capabilities, showcasing its adaptability and generalization skills \cite{hou2024}. Its ability to refactor and annotate code with significant quality improvements further highlights its utility in software engineering \cite{poldrack2023}.

Although excelling in generalization, GPT-4o-mini still faces notable limitations. Its reliance on token limits restricts its ability to process large datasets in a comprehensive way, and exhibits a tendency to hallucinate or produce errors in domain-specific contexts, such as radiology or highly technical programming tasks \cite{liu2023, pordanesh2024}. Like with CodeQwen, we restricted its ability to change code, asking only yes or no questions.
 \section{Encodings}\label{sec_meth}

\begin{wrapfigure}{r}{0.35\textwidth}
  \centering
  \small
\begin{tikzpicture}[
    node distance=2.5cm,
    every node/.style={draw, rectangle, fill=cyan!30, minimum width=1.5cm, minimum height=1cm},
    every edge/.style={->}
]
\node (serialize) {serialize()};
\node (serialize_bytes) [below of=serialize] {bytes (byte[])};
\draw (serialize) -- node[midway, right] {has\_parameter} (serialize_bytes);
\end{tikzpicture}
    \caption{An example TDG snippet. A function $serialize()$ that has a parameter called bytes of type $byte[]$. $serialize()$ is an expression node. $bytes$ is a symbol node. $Has_parameter$ is a merge node.}
    \label{fig:tdg}
\end{wrapfigure}

This section details the three encodings that we considered. The first
two encodings are for the graph model architectures (GCNs and GTNs); the
last encoding is for the text-based LLM models:
\begin{itemize}
\item the type dependency graph (TDG) encoding from prior work (\cref{sec:tdg}),
\item our novel type-system-specific \napast encoding (which is described via the methodology
  to construct a \napast encoding for an arbitrary type system in \cref{sec:napast}), and
\item the text-based encoding via prompting (\cref{sec:finetune}).
\end{itemize}
  
\subsection{Type Dependency Graph}
\label{sec:tdg}

Prior work~\cite{peng2022static} used the TDG encoding for inferring type annotations for programs in dynamically-typed languages such as
Python. The TDG is our baseline encoding for graph models: it has worked for a similar problem in the past.
A TDG has four node types:
1) \emph{symbol nodes}, which represent variables for which the types need to be inferred. Each occurrence of a variable is treated as a unique node to handle scenarios where the variable's type may change at runtime.
2) \emph{expression nodes}, which represent expressions that generate types, such as arithmetic operations, function calls, and list comprehensions.
3) \emph{branch nodes}, which represent branching points in data flow, typically corresponding to conditional statements.
4) \emph{merge nodes}, which represent merging points in data flow, such as the convergence of different branches of execution.
Our implementation of TDGs for Java involved specific adjustments to handle Java-specific syntax and semantics.
For example, we have removed the possibility that a variable's types may change at runtime.
We trained TDGs on a GCN model. A GTN would not confer much benefit since the edges of a TDG have relatively homogeneous semantics. However, the number of types inferred for a pluggable type system is small in comparison to the use case a TDG was designed for. Therefore, the main functionality proposed by these systems is underutilized for predicting pluggable types, motivating our \napast encoding.
An example TDG is shown in \cref{fig:tdg}.

\subsection{Our Novel Graph Encoding: the \napast}\label{sec:napast}

This section describes our novel methodology for building a type-system-specific
program encoding for the graph models (GCN and GTN), which we call a \emph{Name-Augmented, Pruned Abstract
Syntax Tree} (\napast).
Unlike a TDG, this encoding only includes information relevant to the property
being enforced by a particular pluggable type system,
in a form suitable for training a deep-learning model.
The overall methodology for \napast construction is shown
in \cref{fig:overview}.
We explain how we built a \napast encoding for a nullability
type system in detail later, in \cref{sec_train}; this section
gives an overview of the \emph{process} of \napast construction
for an arbitrary pluggable type system.

The base of a \napast is the program's Abstract Syntax
Tree (AST). The first step (\cref{sec:ast-extraction})
to construct a \napast is to augment the AST
with additional nodes that model information traditionally stored
in the AST directly (for example, whether a declaration is \<public> or \<private>),
thereby making that information available to the deep-learning model later.
This is necessary because the internal complexity of the AST data structure 
produced by JavaParser is not visible to the deep-learning model.
\looseness=-1

\newcommand{\yellowboxwidth}{3cm}

\begin{wrapfigure}{r}{0.5\textwidth}
  \centering
  \small
\begin{tikzpicture}
\node[draw, rectangle, fill=yellow, minimum width=\yellowboxwidth, minimum height=1cm] (box1) {1. AST Extraction};
\node[draw, rectangle, fill=yellow, minimum width=\yellowboxwidth, minimum height=1cm, right=of box1, xshift=0.005cm] (box2) {2. Pruning};
\node[draw, rectangle, fill=yellow, minimum width=\yellowboxwidth, minimum height=1cm, below=of box2, xshift=0.005cm] (box3) {3. Name Augmentation};
\node[draw, rectangle, fill=yellow, minimum width=\yellowboxwidth, minimum height=1cm, left=of box3, xshift=0.005cm] (box5) {4. Statement Pruning};
\node[draw, rectangle, fill=yellow, minimum width=\yellowboxwidth, minimum height=1cm, below=of box5, xshift=0.005cm] (box4) {5. Clustering};
\node[draw, rectangle, fill=yellow, minimum width=\yellowboxwidth, minimum height=1cm, below=of box4, xshift=0.005cm] (box7) {6. Training};
\node[draw, rectangle, fill=yellow, minimum width=\yellowboxwidth, minimum height=1cm, right=of box4, xshift=0.005cm] (box8) {7. Prediction};
\node[draw, rectangle, fill=yellow, minimum width=\yellowboxwidth, minimum height=1cm, below=of box8, xshift=0.005cm] (box9) {8. Post-processing};

\draw[->, thick] (box1) -- (box2);
\draw[->, thick] (box2) -- (box3);
\draw[->, thick] (box3) -- (box5);
\draw[->, thick] (box4) -- (box8);
\draw[->, thick] (box4) -- (box7);
\draw[->, thick] (box8) -- (box9);
\draw[->, thick] (box5) -- (box4);
\end{tikzpicture}
\caption{Order of steps in \napast construction.
}
\label{fig:overview}
\end{wrapfigure}
 
The next step in constructing a \napast is 
pruning information
that is not relevant to the property of interest from the graph produced by the previous
step.
We do this by conducting a series of targeted ablation experiments: we iteratively remove particular types
of AST nodes (and all associated edges) from the \napast and retrain a preliminary model using a small dataset
(976 classes, in our instantiation of this process; this number is related to batch sizes and memory limitations on the training machine).

The F1 score was calculated by comparing the predicted nodes with the ground truth using the values in the confusion matrix, namely the harmonic mean of precision and recall. The F1 score on this preliminary model was used to determine which nodes to prune.

Node types whose removal does not negatively impact the F1 score of the preliminary model
are removed from all \napasts. Shrinking the size of the \napast by removing
irrelevant information is critical to the scalability of our approach, since
the model architectures we consider are highly sensitive to the size of the input graph.
There are three pruning phases:
\begin{enumerate}
\item prune nodes that are provably unrelated to the property of interest (\cref{sec:primitives}).
\item use an ablation experiment to prune \emph{node types} that are unrelated to the target property
  (\cref{sec:node-ablation}).
  This pruning step occurs \emph{before} name augmentation when processing the AST for
  a particular class.
\item use an ablation experiment on program statement node types to prune
  \emph{entire subtrees} under each type of node, as well as related
  nodes (\cref{sec:statement-ablation}). This pruning
  step occurs \emph{after} name augmentation when processing the AST for a particular class,
  so we defer discussing it until after we introduce the name-augmentation step.
\end{enumerate}

\noindent
Each pruning phase is theoretically optional, but we found that all three were necessary
to produce \napast graphs that are small enough to realistically train a deep-learning model.
If the pruning steps were not applied, less relevant code would be available to the model in each step (because the model is memory-bound). This makes the model’s performance suffer during prediction:
in our early experiments on a subset of the full dataset, we found that omitting pruning reduced the model's F1 score from 0.95 to 0.78.

The last \napast construction step is 
augmenting the graph with edges encoding program elements with the
same name, which is described in \cref{sec:name-augmented}. The goal of this step is
to encode basic dataflow information in a way that is visible to a deep-learning model.
For example, this stage lets a model determine that a parameter checked against
\<null> in a method should be annotated as \<@Nullable> by connecting the check to
the parameter's declaration.

In this section, we assume the existence of a training dataset of human-annotated
code that contains type qualifiers for the type system of interest. We discuss
whether such datasets exist for real type qualifiers in \cref{ssec_dp}.

\subsubsection{Abstract Syntax Tree Extraction}\label{sec:ast-extraction}

For each class in the training dataset, we first
generate its Abstract Syntax Tree (AST). Traversing the AST, we
generate a graph consisting of nodes and edges.
Unlike HiTyper's Type Dependency Graph~\cite{PengGLGLZL2022}, we have not attempted to build a graph of type dependencies.
We have been able to obtain good results without creating a complex data structure.
We think this is because pluggable type systems are much simpler than the type system of a language.
Each node is represented by a node id, node type and a special field indicating whether the
node contained a type qualifier in the input data (and if, which type qualifier).
Any actual AST nodes for the
type qualifiers generated by the parser are discarded, so that the human-written type qualifiers are
not visible to the model during training. Edges in this graph are
parent-child connections represented as an adjacency list.

The graph contains two types of edges: parent-child, and child-parent.
Representing the internals of each block or statement by a different edge type prevents relevant properties from leaking out into proximate but otherwise unrelated nodes in the AST. On the other hand, we do want statements using the same variables to be proximate in the graph. We achieve this later using the name layer described in \cref{sec:name-augmented}.

We then add additional, useful information that is traditionally stored in the
AST nodes themselves---in particular, modifiers (e.g., \<public>, \<static>, etc.) ---to
the graph's structure, which makes it available to the models during training.
While this may seem like an implementation detail, we mention it here because practical parsers
often store this important information not in the AST itself but in
fields that remain unrepresented in the graph structure.
We represent this information by creating a single node in the graph for each modifier and
connecting the modifiers to the relevant nodes using a multi-hot vector encoding.

\subsubsection{Pruning Phase 1: Remove Provably-Unrelated Information}\label{sec:primitives}

When considering a particular type system, we usually
know that some parts of the program are definitely completely irrelevant to the property
of interest. If we formalize that notion at the AST level, irrelevant parts
of the program can be immediately removed from the AST. For example, a pluggable type system
that tracks whether integers are non-negative (which would be useful as e.g., part of a larger
system for typechecking array accesses) need not consider parts of the program
that do not interact with integers. Therefore, the first pruning stage of \napast construction
is to remove any nodes that are provably irrelevant. For example, primitive types cannot store null values in Java, and so \NullGTN removes them at this stage.
The set of nodes that are irrelevant will
differ when constructing a \napast for each different type system; this pruning stage exists
to allow someone following our methodology to build an inference model for a new type system
to prune anything they know is irrelevant right away, before the more complex pruning stages.

\subsubsection{Pruning Phase 2: Ablation by Node Type}\label{sec:node-ablation}

Like some Java types are irrelevant for predicting pluggable types, so are some AST node types.
For example, MemberValuePair and IntersectionType nodes are irrelevant for predicting non-compound types.
To summarize the AST further, we conduct an ablation study on node types.
For each node type, we drop it and calculate the F1 score on a held-out test set.
We did the same test by dropping random nodes.
We drop node types which score worse than dropping random nodes from the AST. Note that when we prune a node, we connect the node’s children to its parents.

\subsubsection{Name Augmentation}\label{sec:name-augmented}

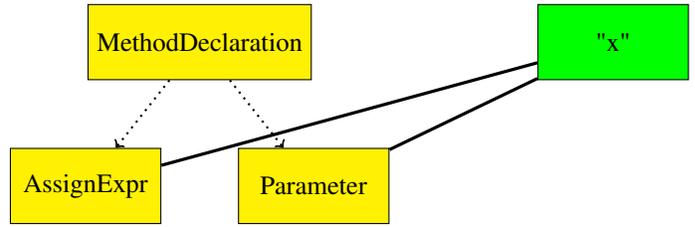
\begin{wrapfigure}{r}{0.5\textwidth}
\centering
\begin{tikzpicture}
\node[draw, rectangle, fill=yellow, minimum width=2cm, minimum height=1cm] (box1) {MethodDeclaration};
\node[draw, rectangle, fill=yellow, minimum width=2cm, minimum height=1cm, below right=of box1.south west, xshift=1cm, yshift=0.1cm] (box2) {Parameter};
\node[draw, rectangle, fill=yellow, minimum width=2cm, minimum height=1cm, below left=of box1.south east, xshift=-1cm, yshift=0.1cm] (box3) {AssignExpr};
\node[draw, rectangle, fill=green, minimum width=2cm, minimum height=1cm, right=of box1, xshift=-0.25cm] (box4) {"x"};

\draw[->, dotted, thick] (box1) -- (box2);
\draw[->, dotted, thick] (box1) -- (box3);
\draw[-, very thick] (box4) -- (box2);
\draw[-, very thick] (box4) -- (box3);
\end{tikzpicture}
\caption{Name Augmentation. This diagram shows the pruned AST as yellow nodes, and a node ("x") in the name layer in green. All the nodes that use the same name are connected by a second edge and node type.}
\label{fig:name-example}
\end{wrapfigure}

To inform the model of the identity of a candidate node, we layer a graph with different
node and edge types on top of the graph derived from the AST
in previous steps. This layer connects the names of the
variables to nodes that make use of those variables,
using different node and edge types. This improves
accuracy by allowing the model to
connect \emph{uses} of names
to their \emph{declarations}.
Since type annotations need to be placed on declarations,
this step is critical: the model must know that a particular test or assignment is related to a specific declaration.
For example, \cref{fig:name-example} shows an example
for a variable named "x", which is
connected as a separate graph layered over the pruned AST.
The names are connected using a different edge type. The second set of edges are stored in a different adjacency matrix.

This layer can be viewed as encoding a simplistic dataflow analysis,
or alternately as enabling the model to reason about basic dataflow.
We chose to add this graph layer at the encoding stage rather than
running a dataflow analysis directly because it was simpler, and
in our experiments it appears to work well in practice. Future work
could investigate making dataflow information available to model
directly as an alternative to our name augmentation strategy.

\subsubsection{Pruning Phase 3: Ablation of Statement Subtrees}\label{sec:statement-ablation}

Even after the pruning phases described above, after name
augmentation the resulting graph is prohibitively large to train
the graph-based models in
\cref{sec:model}. So, we need another pruning strategy.
Our key insight at this stage is that many statements in the program
are irrelevant to the property of interest. For example, parts of the program that
do not include any null literals are unlikely to be related to
nullability. However, dropping \emph{all} such statements hurts model performance (as measure by F1
score) too much, so we performed an ablation study to determine
which kinds of statements are safe to prune.
Our goal at this stage is to prune statements, their corresponding
subtrees, and any other related nodes. We use the name-augmentation
layer added to the graph in \cref{sec:name-augmented} to determine
which non-subtree nodes are related to a statement: when we prune
a statement and its subtree, we also prune any node that is directly
connected to the statement's subtree via a name layer.

\subsubsection{Clustering}\label{sec:clustering}
In machine learning projects, one typically cleans the dataset by pruning outliers. Instead of trying to visualize the different kinds of graphs so that we can prune outliers by hand,
we clustered the dataset using k-means clustering and trained a separate model for each cluster.
We chose a value of k=5 by minimum within-cluster variation with the elbow method~\cite{cui2020introduction}.
This method raised the overall F1 score by approximately 0.5, so it is critically-important
to overall model performance.

\subsubsection{Prediction}\label{sec:predict}
A graph-based model typically makes a prediction for a single input graph (in our case, a \napast
representing a single class) at a time. However, program elements defined in one
class are sometimes used in other classes in ways that impact their nullability.
To capture this phenomenon, we employ a \emph{conjoined prediction}
strategy: for each pair of classes in a target project, we predict the nullability
of all of their elements together, using JavaParser's symbol solver to connect
the name layers of the \napasts if they have any elements in common.
The final prediction for an element's nullability is the average of the predictions
across all pairs of classes.
This technique increases the consistency of the model's prediction across
classes: conjoined prediction helps when an element is used in more than one
class, but makes no difference for elements that are not used in the other class
for which predictions are being made concurrently. In other words, this conjoined
prediction strategy allows the model to reason (in a limited way) about inter-class
dataflow.

\subsubsection{Post-processing}\label{sec:post}
Deep-learning methods are good for finding recommendations through
hidden patterns in the data. Unfortunately, their outputs are not
guaranteed to be internally consistent.
Post-processing
steps remove obvious inconsistencies: they certainly
do not guarantee that the model's outputs will be sensible.
They should, however, be conservative in a sense: the annotations
they add or remove from the model's output must be guaranteed to make the
resulting annotation set for the whole program more consistent.

\subsubsection{Other types}
The NaP-AST encoding can be adapted with relevant changes to any type system with finite types. For example, consider a type system for the signedness of integers. The relevant node types, etc., would almost certainly differ from the things that matter for null safety (e.g., null literals might not matter, but addition or subtraction nodes might matter a lot), so the process that we describe in \cref{sec_meth} would need to be repeated. We separated the description of the process in \cref{sec_meth} from the instantiation of that process for null safety in \cref{sec_train} to facilitate repeating our approach in the future.

\subsection{Text Encoding}\label{sec:finetune}

To have LLMs place annotations, we could not provide it with complete source files,
because many were too large.
We extracted segments that fit within a conservative ~2 K-token budget to keep inference consistent across models and reduce nonsensical outputs,
even though some models (e.g., CodeQwen 1.5-7B-Chat) support much larger contexts (up to 64 K).

To avoid hallucinations that change code structure, we restrict the output of LLMs to yes or no, using them as classifiers.
First, the script extracts potential candidates for @Nullable annotation from the Java file by scanning its lines with two regular expressions, field declarations  and method signatures. The latter captures both return types and parameter lists. For each match, a unique identifier and their corresponding elements are stored in a dictionary for further processing.
Next, a prompt is constructed consisting of a concise system message ("You are a helpful assistant.") followed by a detailed user message. The user message provides the full Java source code and lists each extracted candidate element, prefixed by its unique key. The LLM is then asked to decide for each candidate whether it should be annotated with @Nullable. The script explicitly instructs the LLM to return a strictly formatted JSON object that maps each candidate key to either "yes" or "no." For any candidate marked with "yes," it inserts the @Nullable annotation in the appropriate location, whether at a field, method return, or parameter.
Without these guardrails, we found that the LLMs break most code segments trying to make syntactic changes---for example, by adding
or removing import statements unnecessarily.
Our prompting code's rules try to include ending braces and complete loops, methods or files where possible to help reduce the impact of these problems.

 \section{Dataset}\label{ssec_dp}

To evaluate the encodings in \cref{sec_meth} and to answer
our other two research questions---i.e., which model is best suited to the task
and how much data is actually required---we performed a feasibility study on
the most popular pluggable type system: preventing null-pointer dereferences
in Java using a \<@Nullable> annotation. Multiple typecheckers for preventing
this problem exist and have been deployed in industry, including
Uber's NullAway~\cite{BanerjeeCS2019}, Meta's NullSafe~\cite{PianykhZL2022},
and the Checker Framework's Nullness Checker~\cite{PapiACPE2008,DietlDEMS2011}.

The first challenge in predicting nullability annotations is collecting training data.
Each nullability checker has its own set of annotations.
Further, there are annotations provided by frameworks (e.g., the JSR 305 nullability annotations
in the JDK, or those from Android) without an associated checking tool.
This fragmentation in the ecosystem means that for any single nullability checker or annotation,
there is not enough training data to build an effective model. This mirrors the
general situation for pluggable typecheckers, wherein most checkers are used by only a 
small number of projects, leading to too little training data when considering whether
training a model is practical.

Our key insight in dataset preparation is that the various nullability tools all use
similar definitions of \<@Nullable>: it universally means ``this element may
or may not be null.'' Since the most widely-used pluggable typecheckers use \<@NonNull>
as the default type for unannotated code (e.g., both NullAway~\cite{BanerjeeCS2019} and the Checker
Framework~\cite{PapiACPE2008,DietlDEMS2011} use this defaulting scheme),
bringing a program into compliance with a typechecker usually requires a human to write
mostly or only \<@Nullable> annotations. Consequently, we can restrict a model's goal to the prediction of only \<@Nullable> annotations:
we do not need to directly predict \<@NonNull> or any other, more complex annotations.
In turn, the need to only predict \<@Nullable> means that
we can \emph{combine} training data that uses different \<@Nullable> annotations, as long
as those annotations have the same semantics.
We used every annotation in the comprehensive list of \<@Nullable> annotations with the same semantics
in the Checker Framework's manual~\cite{cf-nullness-list}.

We used the search feature of Sourcegraph's CLI
tool~\cite{b8} to collect Java classes that import \<@Nullable> annotations. We searched for \<@Nullable> annotations by themselves. However, this approach led to a dataset where JetBrains projects were overrepresented. We think this happened because Jetbrains has the most popular projects that use the \<@Nullable> annotation. We switched to searching for a variety of Nullable imports to gather a diverse dataset.
We do not require code in our training set be checked by some nullability
tool: instead, we try to model where a human would write a nullability annotation.
E.g., \<@Nullable> annotations that a human wrote as documentation are also useful
as training data.

Some classes, despite importing a nullability annotation, did not
actually contain any such annotations. After selecting only
those classes that did, we removed very small and very large
classes.
We define “very small” classes as those with fewer than 300 AST nodes. There are 11,019 such classes in the dataset. Based on manual review, we observed that these classes lacked nontrivial logic (besides getters, setters and annotated fields). Our early experiments showed that retaining them in the training set skews the model towards annotating every annotatable location.
We define “very large” classes as those with more than 8,000 nodes after pruning. We decided to remove these because we wanted the model to see the complete code of each class during training. 8,000 nodes was the empirical limit imposed by the memory capacity of the training machine.
After all of these removals, we were left with 32,370 classes;
we removed 60,292 classes in total.
Note that despite excluding both categories during training, the models  operate on all classes regardless of size when deployed. The NullAway benchmark (section 7.1) contains 1641 “too small” classes and 24 “too large” classes that would have been removed during training. Our results there show that removing these classes during training does not cause problems during deployment.
Each class in this dataset contains one or more \<@Nullable> annotations.
There are 217,922 \<@Nullable> annotations in these classes in total.
 \section{\napasts for Nullability}\label{sec_train}

With the dataset in \cref{ssec_dp}, we trained graph-based models (GCN and GTN)
using our \napast encoding method (\cref{sec:napast}).
This section describes the \napast construction process; we call the resulting models
NullGCN and \NullGTN.

\subsection{\napast Pruning}

In this section, we describe the specific pruning strategies that we used for each
pruning phase described in \cref{sec:napast}.

\subsubsection{Phase 1: Pruning by Proof}

Since primitive types cannot store null values in Java, we exclude them from
the nodes that the model is trained on. Nullability is relevant to most of the program,
but we expect that for other pluggable type systems that target rarer problems,
this step could prune dramatically-larger parts of the program. For example,
an instantiation of our methodology for a type system that detects resource leaks
could remove all parts of the program that definitely do not manipulate resources
(i.e., most of the program).

\subsubsection{Phase 2: Pruning by Node Type}

\begin{figure}[t!]
 \begin{tikzpicture}
 \begin{axis} [
   xlabel={Node Type},
   ylabel={F1 after dropping},
   grid=major,
   symbolic x coords={InstanceOfExpr, ArrayInitializerExpr, TypeExpr, MethodCallExpr, ForStmt, VariableDeclarator, UnaryExpr, VarType, NullLiteralExpr, NormalAnnotationExpr, IfStmt, LambdaExpr, ArrayCreationLevel, EnclosedExpr, BinaryExpr, ImportDeclaration, ExplCtorInvocStmt, ReturnStmt, TryStmt, ForEachStmt, IntegerLiteralExpr, WildcardType, ConstructorDeclaration, LineComment, ConditionalExpr, ObjectCreationExpr, ArrayType, <random>, ArrayCreationExpr, ArrayAccessExpr, CharLiteralExpr, IntersectionType, AssignExpr, SingleMemAnnoExpr, MemberValuePair, ClassExpr, Parameter, EnumDeclaration, BlockStmt, TypeParameter, ClassOrInterfaceDecl, Modifier, ThisExpr, MethodDeclaration},
   xtick=data,
   xticklabel style={rotate=90, anchor=north east},
   xticklabel={
       \ifthenelse{\equal{\tick}{<random>}}{\textcolor{red}{\tick}}{\tick}
   },
   ymin=0.6, width=\columnwidth, height=6cm,
   tick label style={font=\tiny},
   xlabel style={yshift=-0.1cm},
 ]

 \addplot[red,only marks,mark=square*] coordinates {
   (InstanceOfExpr,0.9492)
(ArrayInitializerExpr,0.9441)
(TypeExpr,0.939)
(MethodCallExpr,0.9351)
(ForStmt,0.9305)
(VariableDeclarator,0.9219)
(UnaryExpr,0.9151)
(VarType,0.9132)
(NullLiteralExpr,0.9058)
(NormalAnnotationExpr,0.9035)
(IfStmt,0.8986)
(LambdaExpr,0.8973)
(ArrayCreationLevel,0.8961)
(EnclosedExpr,0.8959)
(BinaryExpr,0.8957)
(ImportDeclaration,0.8927)
(ExplCtorInvocStmt,0.892)
(ReturnStmt,0.8917)
(TryStmt,0.8862)
(ForEachStmt,0.8838)
(IntegerLiteralExpr,0.8775)
(WildcardType,0.8748)
(ConstructorDeclaration,0.8721)
(LineComment,0.8712)
(ConditionalExpr,0.8666)
(ObjectCreationExpr,0.8656)
(ArrayType,0.8566)
(<random>,0.8561)
(ArrayCreationExpr,0.8559)
(ArrayAccessExpr,0.8473)
(CharLiteralExpr,0.8438)
(IntersectionType,0.8436)
(AssignExpr,0.8354)
(SingleMemAnnoExpr,0.8294)
(MemberValuePair,0.8069)
(ClassExpr,0.8032)
(Parameter,0.8018)
(EnumDeclaration,0.7828)
(BlockStmt,0.7732)
(TypeParameter,0.7722)
(ClassOrInterfaceDecl,0.7699)
(Modifier,0.7604)
(ThisExpr,0.713)
(MethodDeclaration,0.666)
 };
\end{axis}
 \end{tikzpicture}
 \caption{
   Ablation study to decide which node types can be dropped without affecting the F1 score. Only
   node types that performed worse than dropping random nodes (in red) were retained in later models.
 }
 \label{fig_ab1}
\end{figure}
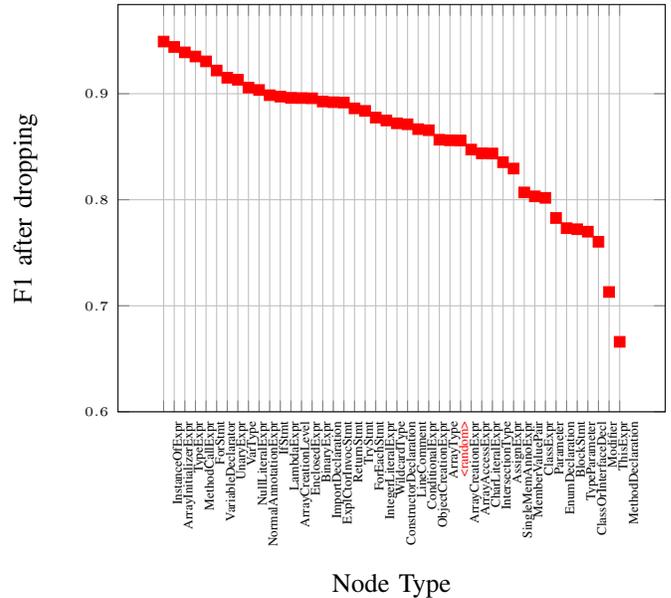

First, we conducted an ablation study over node types.
We dropped each node type and calculated the F1
score of a preliminary model on held-out testing data.The results are plotted in \cref{fig_ab1} (test runs were repeated 50 times each to control for the statistical nature of the models).
Note ``<random>'' in red in \cref{fig_ab1}: this represents dropping nodes at random from
the graph, without regard to their type.
We discarded all nodes of the types that appear to the left of ``<random>'' in
\cref{fig_ab1}: that is, node types whose removal damages the model's predictive
ability less than removing random nodes. We connect the children of
dropped nodes to their parents. \looseness=-1

For example,
since humans often write \<@Nullable>
annotations on method declarations (to indicate nullable return types), dropping them
dramatically reduces the F1 score to about 0.66 .
By contrast, node types like \<MemberValuePair> and
\<IntersectionType> seem to have no measurable impact on the performance
of the model,
which makes sense: value pairs are unrelated to reasoning about nullness.
As another example, \<IntersectionType> nodes indicate a generic
type construct---which is entirely orthogonal to nullness. 

Dropping some node types surprisingly drops the F1 score
(if \<ClassOrInterfaceDeclaration> is dropped, for example,
the F1 score falls to 0.76).
We determined via manual investigation that
nodes of this type help group the methods of nested classes, so even though they are rarely
or never nullable
themselves, retaining them is important to model performance.
Another node type that helps group nodes together is \<BlockStmt>;
dropping them reduces the F1 score to 0.77.

\subsubsection{Phase 3: Pruning By Subtree}

\begin{figure}[t!]
 \begin{tikzpicture}
 \begin{axis} [
   xlabel={Node Type},
   ylabel={F1 after dropping},
   grid=major,
   symbolic x coords={ForStmt, BlockStmt, LocalClassDeclStmt, SynchronizedStmt, ForEachStmt, CatchClause, WhileStmt, LabeledStmt, AssertStmt, ThrowStmt, <random>, ExplCtorInvocStmt, ReturnStmt, IfStmt, ContinueStmt, SwitchStmt, TryStmt, ExpressionStmt},
   xtick=data,
   xticklabel style={rotate=45, anchor=north east},
   xticklabel={
       \ifthenelse{\equal{\tick}{<random>}}{\textcolor{red}{\tick}}{\tick}
   },
   width=\columnwidth, height=6cm,
   tick label style={font=\scriptsize},
   xlabel style={yshift=-0.1cm},
 ]

 \addplot[red,only marks,mark=square*] coordinates {
   (ForStmt,0.9489)
(BlockStmt,0.9234)
(LocalClassDeclStmt,0.9226)
(SynchronizedStmt,0.9189)
(ForEachStmt,0.906)
(CatchClause,0.898)
(WhileStmt,0.88)
(LabeledStmt,0.8762)
(AssertStmt,0.8709)
(ThrowStmt,0.863)
(<random>,0.861)
(ExplCtorInvocStmt,0.8538)
(ReturnStmt,0.8479)
(IfStmt,0.8471)
(ContinueStmt,0.8388)
(SwitchStmt,0.8264)
(TryStmt,0.819)
(ExpressionStmt,0.7743)
 };
 \end{axis}
 \end{tikzpicture}
 \caption{Results of the ablation study in \cref{sec:statement-ablation}
   by statement type. Subtrees of nodes of types
   to the left of ``<random>'', and nodes connected to those subtrees by the name
   layer, are pruned from the \napast.}
 \label{fig_ab2}
\end{figure}
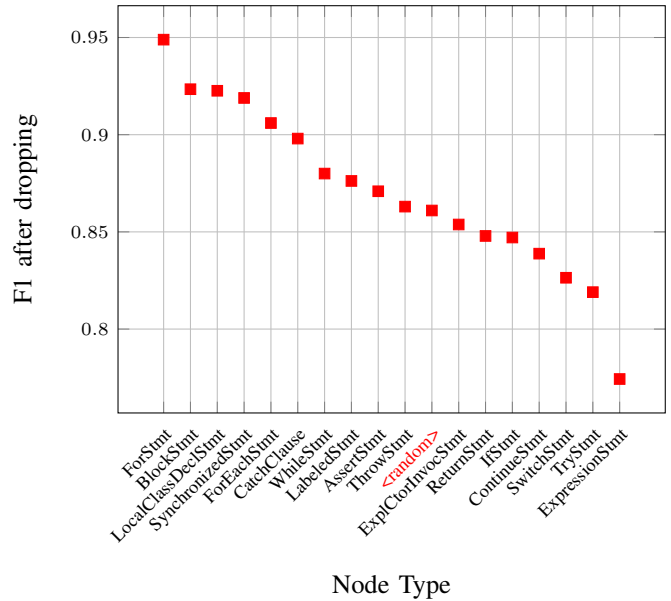

Using a preliminary model, we conducted an ablation study
to determine which kinds of statement nodes we can remove,
if the corresponding subtree does not contain a \<NullLiteralExpr> node.
In the ablation study, we pruned all such subtrees under a particular
kind of statement node (and all nodes connected to that subtree via
a name layer), for each kind of statement. The resulting
F1 scores for the preliminary model appear in \cref{fig_ab2}.
Our \napast construction
procedure prunes the subtrees and name-connected nodes for the
types of statement nodes that appear to the left of the ``<random>''
entry (in red) in \cref{fig_ab2}: that is, those statement types that are
less useful to the model than a random statement. For example,
a nullability \napast will not contain a \<for> loop (\<ForStmt> in \cref{fig_ab2})
nor nodes for any of the variables manipulated in that loop unless
there is a null literal within the loop. However, all
expression statements (not within the subtree of another pruned
statement) will be preserved, whether or not the \<ExpressionStmt> node
for the statement has a null literal in one of its
subtrees---removing expression statements damages model performance too much.

\subsubsection{Post-processing}
Four nullability-specific
post-processing steps ensure internal consistency.

\paragraph{Returning Nullable Fields}\label{sec:nullable-field-return}
Our first post-processing step is a simplistic dataflow analysis:
when a method returns a field that the model has marked as \<@Nullable>,
the method's return type should be \<@Nullable>, too. This post-processing
step is especially important for getters, which the models otherwise rarely
predict are nullable.

\paragraph{Consistency between Arguments and Parameters}\label{sec:nullable-param-arg}
Our next post-processing step is similar in spirit: at a call-site
where an argument is a read of a field that the model predicted is \<@Nullable>,
we change the corresponding parameter to \<@Nullable> in the method declaration.
We have observed, qualitatively, that these models
tend to do a poor job predicting whether parameters are nullable, relative to
other kinds of program elements like fields; we conjecture that the reason for
this behavior is that nullable fields are often directly assigned null, but
many methods with nullable parameters are never directly called with null literal
arguments.

\paragraph{Stream Parameters}\label{sec:stream-parameters}
Since the models use piece-wise boundaries in a continuous, higher
dimensional space, their impressionistic understanding of Java syntax
occasionally puts annotations where they cause compilation
errors. For example, a model might place a \<@Nullable> annotation in a stream
parameter, where Java's syntax does not permit one. We observed
this in practice, so we added a post-processing step to remove these \<@Nullable>
annotations.\looseness=-1

\begin{wraptable}{l}{4.5cm}
  \caption{GTN hyperparameters}
  \centering
  \begin{tabular}{c | c}
    \hline
    Hyperparameter & Value\\
    \hline
    GT Layers &   5\\
    FGTN Layers &   2\\
    Epochs   & 10/batch\\
    Learning Rate &   0.01\\
    Channel Agg. & Mean\\
    \# of Channels & 2\\
    Non-local weight & -2\\
    k-Hop & 9\\
    \hline
  \end{tabular}
  \label{tab_hyp} 
\end{wraptable}

\paragraph{Inheritance}\label{sec:superclass}
A method's inferred parameter or return types in the same inheritance hierarchy
may not match. We remove \<@Nullable> annotations in
superclasses and interfaces when no such annotation was inferred
in the subclass or implementing class, respectively.

\section{Training}\label{sec:training}
We split the nodes with \<@Nullable> annotations into 60\% training, 20\%
validation, and 20\% test datasets. To create a balanced dataset, we
had to combine each set with an equal number of nodes without
\<@Nullable> annotations. To do this, we took the annotated nodes in the
dataset and subjected them to the 60-20-20 split. We combined the
training and validation sets with unannotated nodes of equal number from the same classes,
and put all remaining nodes in the test set.

We trained the GTN on a cluster with Intel
Xeon E5-2630-v4 CPUs (2.2 GHz), 2 GPUs and 50 GB RAM in the Lochness HPC.
Since the adjacency matrices that we use to represent the AST
are sparse, they take up a lot of space on the GPU.
So, we were only able to train a sample consisting of 1.15\% of our dataset at a time.
We thus chose batch training, which allows
us to train the GTN on the full dataset over a longer period of time.
We fixed the input size to 8000 nodes,
and only trained on as many classes at a time as fit into that space.
The GCN's training (and all the experiments in \cref{sec_expr})
were run on an AWS g4dn.xlarge instance and a Deep Learning PyTorch AMI. We trained NullGCN using PyTorch and the PyTorch Geometric library.
The GCN uses two convolutional layers, dropout regularization, negative log-likelihood loss function and the Adam optimizer.

\subsubsection{Hyperparameters}
After some trial and error, we used a simple 2 layer GCN with batching.
The GTN's trained hyperparameters are in \cref{tab_hyp}. These hyperparameters
were chosen by manual grid search. For this dataset, non-local weight of -2 performs well independent of other variables. Since we trained a large dataset in batches, we chose a low initial learning rate and used weight decay throughout the training process to avoid overfitting.

\subsection{Thresholding during Deployment}\label{sec:threshold}

The models are trained on an equal number of nodes that are nullable and
non-null, and all the training data was sourced from classes that contained
at least one nullable member. Their predictions reflect this: averaged over many pieces
of code, the average nullability prediction is about half nullable
elements and half non-null elements.
In real code, however, there are many
more non-null elements than nullable elements: in fact, real code contains
many classes that have no nullable members at all!
To account for this, when applying each model, we use a \emph{thresholding}
strategy: we only actually annotate elements as nullable
when the model is very confident in that element's nullability.
In our experiments with real projects in \cref{sec:nullaway-eval}, we applied
a threshold of 90\% confidence.
In a deployment of one of these models on a previously-unannotated codebase
(which is our envisioned deployment scenario), we recommend using
a similar threshold.
 \section{Experiments and Results}\label{sec_expr}

We address the research questions from \cref{sec:intro}:
\begin{researchquestions}
\item What kind of machine-learning model architecture is suitable for inferring pluggable types, and how does that model differ from
  those in prior type inference works for other languages?
\item How much data is needed to make that model effective at inferring qualifiers in practice?
\end{researchquestions}
\noindent We performed two experiments. The first is a large-scale evaluation of each model's ability to infer
ground-truth, human-written annotations on real open-source projects from a previous study
of a nullability checker (\cref{sec:nullaway-eval}) to answer
\textbf{RQ1}.
The second experiment (\cref{ssec:rq3}) evaluated how much training data
the best model from \cref{sec:nullaway-eval} required for good performance, to answer \textbf{RQ2}.
All of our code, dataset, and experimental scripts are open-source and
available~\cite{our-data}.

\subsection{NullAway Benchmark Experiment}\label{sec:nullaway-eval}
We ran each model on the open-source benchmark from the
NullAway paper~\cite{BanerjeeCS2019}.
This benchmark (from their artifact~\cite{nullaway-artifact})
is a collection of 18 projects from GitHub; each was
annotated with nullability annotations when the artifact was published.
The artifact also includes an evaluation tool to run NullAway on these projects.
Two of these repositories no longer exist; we could build the project and run NullAway on 11 of the remaining 16;
two more projects
no longer use \<@Nullable> annotations (in the branches pulled by the artifact's
evaluation tool). We consider the remaining 9 projects that we can typecheck using NullAway;
however, we note that of these projects, 2 of the 9 do not typecheck with no
warnings, even with the human-written annotations. We assume that this is because
of either changes in the projects or changes in NullAway itself.

In the remaining projects, 142 out of the 594 classes (\textasciitilde 23\%) have a \<@Nullable> annotation.
This is a typical ratio; as another point of comparison, consider the eureka and spring-framework projects (both of which are outside the dataset but are annotated for NullAway). Eureka~\footnote{https://github.com/Netflix/eureka} has 21/408 annotated classes (\textasciitilde 5\%); spring-framework~\footnote{https://github.com/spring-projects/spring-framework} has 3016/8800 (\textasciitilde 34\%). These projects are examples of human-annotated projects near the extremes of the distribution. 

We checked each benchmark project with NullAway in three configurations:
1) with the human-written annotations that came with the projects,
2) after removing the human-written annotations from each project,
and 3) after using each model to re-annotate the code.
We measured two proxies
for human annotation effort: each model's precision/recall of the human-written annotations
(which are erased before passing the projects to each model) and the
reduction in the warnings from NullAway using our annotations
versus the unannotated code. The results appear in \cref{tab_nul}.

\begin{table*}[!t]
  \caption{NullAway benchmark results. ``War.'' means ``Warnings.'' ``Red. \%'' means ``reduction percentage'': that is, the reduction in warnings achieved by each model vs. the unannotated code.
  }
  \centering
  \resizebox{\textwidth}{!}{
    \begin{tabular}{c | c | c c c c | c c c c | c c c c | c c c c | c c c c}
      & \multicolumn{1}{c|}{Unannotated} & \multicolumn{4}{c|}{NullGCN} & \multicolumn{4}{c|}{TDG GCN} & \multicolumn{4}{c|}{NullGTN} & \multicolumn{4}{c|}{GPT-4} & \multicolumn{4}{c}{CodeQwen} \\
      Project & War. & War. & Red. \% & Prec. & Rec. & War. & Red. \% & Prec. & Rec. & War. & Red. \% & Prec. & Rec. & War. & Red. \% & Prec. & Rec. & War. & Red. \% & Prec. & Rec. \\
      \hline
ColdSnap & 256 & 2 & 99\% & .06 & .2 & 256 & 0\% & .72 & .28 & 2 & 99\% & .15 & .8 & 0 & 100\% & .65 & .28 & 2 & 99\% & 1 & .26 \\
QRContact & 0 & 0 & - & 0 & 0 & 0 & 0\% & .66 & 1 & 0 & - & 0 & 0 & 0 & - & 1 & 1 & 0 & - & 0 & 0 \\
jib & 101 & 122 & -20\% & .62 & .24 & 101 & 0\% & .9 & .33 & 101 & 0\% & .55 & .68 & 166 & -64\% & .82 & .33 & 101 & 0\% & .9 & .28 \\
meal-planner & 1 & 11 & -1000\% & .07 & .33 & 1 & 0\% & 1 & .66 & 1 & 0\% & .2 & .33 & 9 & -800\% & .25 & .66 & 1 & 0 & 1 & .66 \\
uLeak & 12 & 6 & 50\% & .14 & .4 & 12 & 0\% & 0 & 0 & 8 & 33\% & .41 & 1 & 6 & 50\% & 1 & .2 & 12 & 0\% & 0 & 0 \\
AutoDispose & 2 & 30 & -1400\% & 0 & 0 & 1 & 50\% & 0 & 0 & 2 & 0\% & .01 & .11 & 30 & -1400\% & 0 & 0 & 24 & -1100\% & 0 & 0 \\
picasso & 0 & 0 & - & .61 & .39 & 0 & 0\% & .28 & .8 & 0 & - & .68 & .73 & 3 & - & .66 & .02 & 0 & - & 0 & 0 \\
skaffold-tools-for-java & 1 & 1 & 0\% & 1 & .71 & 1 & 0 & 0 & 0 & 1 & 0\% & .62 & .71 & 2 & -100\% & 0 & 0 & 1 & 0\% & 0 & 0 \\
butterknife & 4 & 4 & 0\% & 0 & 0 & 4 & 0\% & 0 & 0 & 0 & 100\% & 1 & 1 & 4 & 0\% & 0 & 0 & 4 & 0\% & 0 & 0 \\
      \hline
\textbf{Totals} & 377 & 176 & 53\% & .31 & .27 & 376 & 0\% & .57 & .36 & 115 & 69\% & .39 & .69 & 220 & 41\% & .76 & .25 & 145 & 61\% & .91 & .21 \\
    \end{tabular}
  }
  \label{tab_nul} 
\end{table*}
 
As an example of the the task that the models are being asked to do,
consider the following line: \vspace{0.1in}
\begin{lstlisting}
private static LeakInstanceTracker trackerSingleton = null;
\end{lstlisting}
\vspace{0.1in}
A model should infer that this field is \<@Nullable>, and such an annotation is present in the ground-truth human-annotated code;
without an \<@Nullable> annotation, NullAway complains that \<trackerSingleton> is supposed to be non-null and yet it is assigned to null.
Adding the annotation automatically would reduce the human effort to adapt this codebase to using NullAway, which is why we use
warning reduction as a proxy for human effort.

\subsection{RQ1}\label{ssec:rq2}
Our experiments compared the following approaches:
\begin{itemize}
\item Graph Convolutional Networks (GCN)~\cite{zhang2019graph} is the common baseline model for node prediction tasks.
  We trained GCN models on the \napast representation derived in \cref{sec_train} (``NullGCN'').
\item The TDG encoding from prior work~\cite{peng2022static} (``TDG GCN'').
\item A Graph Transformer Network (GTN)~\cite{yun2019graph} is a model that is mathematically guaranteed to perform at least as well as a GCN, but extends its capabilities for graphs with multiple edge types. Our GTN model (``NullGTN'') is also trained on our \napast encoding; the TDG encoding only has one edge type, so it gains no benefit from the GTN over the GCN.
\item Large Language Models (LLMs) like CodeQwen 1.5-7B-Chat and GPT-4 are a good baseline for coding tasks.
\end{itemize}

\paragraph{NullGCN}
On the NullAway benchmark, NullGCN yields 176 warnings vs 377 unannotated (53\% reduction), with precision 0.31 / recall 0.27. While totals improve, the model regresses on some small projects (e.g., meal-planner 1 to 11, -1000\%; AutoDispose 2 to 30, -1400\%), indicating that misplaced annotations can introduce new warnings even as others are removed. While this is
better than no inference at all,
this result suggests to us that the hypergraph structure of the GTN (which does much better)
is important to modeling where humans write annotations:
the same graph is not as useful if the edges are not known to be of different types.

\paragraph{TDG GCN}
TDG GCN achieves precision 0.57 / recall 0.39 but yields ~0\% warning reduction (376 vs 377).
The inferences placed by this model were not accurate enough to reduce warnings on the projects.
We hypothesize that the TDG chains are too disconnected to place annotations in a consistent manner.
We speculate the TDG chains are better suited for use as an additional layer in the NaP-AST rather than an independent encoding.
However, because the TDG encoding only has one edge type, it cannot benefit from the GTN model, and it is dramatically outperformed by the \napast plus GTN combination.
This shows that the dataflow hints in our name layer are critical for inferring pluggable types.

\paragraph{NullGTN}\label{ssec:gtn}
We built our GTN model (\NullGTN) by modifying a fork of the original code~\cite{fastgtn-code}.
It takes an average of about 16 hours to train the model. The largest input
source code we evaluated (6,470 non-comment, non-blank lines) took about 5 minutes for inference.
\NullGTN achieved an F1 score of .95 on the test set during training.
On the NullAway Benchmark, \NullGTN discovered 69\% of the human-written type qualifiers---most of them---with a precision of 39\%.
However, this only eliminated 69\% of the NullAway warnings that a
human would need to resolve to adapt these benchmarks to use NullAway---a large
amount, but far from all of the warnings.
One source of \NullGTN's inability to reduce warnings further is imprecision,
which leads to extra annotations, which leads to extra warnings.
Another cause of low warning reduction for \NullGTN is interactions
with libraries. \NullGTN has no model of what libraries do, and unlike an LLM
it cannot guess based on e.g., the names of library methods. The jib
project (where \NullGTN fails to remove many warnings despite good recall)
uses libraries extensively. It is a modular system with many APIs and plugins,
which the model was unaware of.
Overall, these results are encouraging: \NullGTN makes a good start on
bringing these benchmarks into compliance with NullAway, and finds a majority (69\%)
of the annotations that the developers themselves keep in their codebases.

\begin{wrapfigure}{r}{0.5\textwidth}
\begin{lstlisting}[language=Java]
class LeakTracker {
  private WeakReference<Object> gcIndicatorWeakReference;

  LeakTracker(Class<?> eventClass, WeakReferenceGenerator g, String name, LeakType leakType) {
    this.eventClass = eventClass;
    this.weakReferenceGenerator = g;
    this.name = name;
    this.leakType = leakType;
  }
}
\end{lstlisting}
\caption{Simplified example of code that CodeQwen fails to annotate correctly from uLeak's \<LeakTracker.java>.}
\label{fig:uLeakQwenExample}
\end{wrapfigure}

\paragraph{GPT-4}
To evaluate GPT-4’s suitability for pluggable type inference, we tested its performance on the same dataset used for our NaP-AST-based models. GPT-4 demonstrated moderate precision (76\%) and recall (25\%) compared to the graph-based models, indicating that while it can correctly annotate some types, it struggles to identify the majority of annotation sites. Despite slightly higher recall than CodeQwen, GPT-4’s more aggressive edits occasionally introduce warnings (e.g., picasso 0 to 3), which explains its lower net reduction.

GPT-4’s strength lies in its general coding abilities, but its lack of domain-specific understanding of pluggable type systems hinders its performance. Unlike NullGTN, GPT-4 does not incorporate graph-based dataflow hints, which are crucial for accurate type inference.

\paragraph{CodeQwen 1.5-7B-Chat}
On the NullAway benchmark (Table~\ref{tab_nul}), CodeQwen achieved 91\% precision and 21\% recall, reducing warnings by 61\% overall. However, its edits can increase warnings on some projects (e.g., AutoDispose: 2 to 24 warnings), consistent with occasional unstable edits even at temperature 0.

The precision value is very high because it places relatively few annotations compared to the other models. This is because of CodeQwen's greater conservatism in making changes. For example, in
the code example in \cref{fig:uLeakQwenExample},
the \<gcIndicatorWeakReference> field should be nullable because it is not initialized by the constructor. But since LLMs are trained to reason based on present tokens instead of absent ones, CodeQwen misses it.

\paragraph{Answer to RQ1}
Across the 9-project benchmark (Table~\ref{tab_nul}), NullGTN attains the highest recall (0.69) and warning reduction (69\%) at moderate precision (0.39). CodeQwen has the highest precision (0.91) but lower recall (0.21) and reduces warnings by 61\% overall.
The other models and encodings all perform strictly worse than one of these two models.
Graph models—especially NaP-AST + GTN—deliver the best overall downstream performance (highest recall and largest warning reduction). LLMs achieve very high precision (e.g., CodeQwen 91\%) but lower recall. This suggests that a graph encoding is the right choice for type inference,
which is in line with prior work on non-pluggable type inference~\cite{peng2023statistical,wei2023typet5,peng2023generative,PengGLGLZL2022,mir2022type4py,cui2021pyinfer}.

\begin{wrapfigure}{r}{0.5\textwidth}
 \begin{tikzpicture}
 \begin{axis} [
   xlabel={Training percentage},
   ylabel={Warnings},
   grid=major,
   symbolic x coords={10, 20, 30, 40, 50, 60, 70, 80, 90, 100},
   xtick=data,
   xticklabel style={rotate=45, anchor=north east},
   xticklabel={
       \ifthenelse{\equal{\tick}{<random>}}{\textcolor{red}{\tick}}{\tick}
   },
   width=0.5\textwidth, height=6cm,
   tick label style={font=\scriptsize},
   xlabel style={yshift=-0.2cm},
   ymin=0,  ]

 \addplot[red,only marks,mark=square*] coordinates {
   (10,655)
(20,720)
(30,712)
(40,611)
(50,374)
(60,224)
(70,286)
(80,724)
(90,686)
(100,717)
 };
 \end{axis}
 \end{tikzpicture}
 \caption{Warnings by percentage of data trained.}
 \label{fig_perc}
\end{wrapfigure}
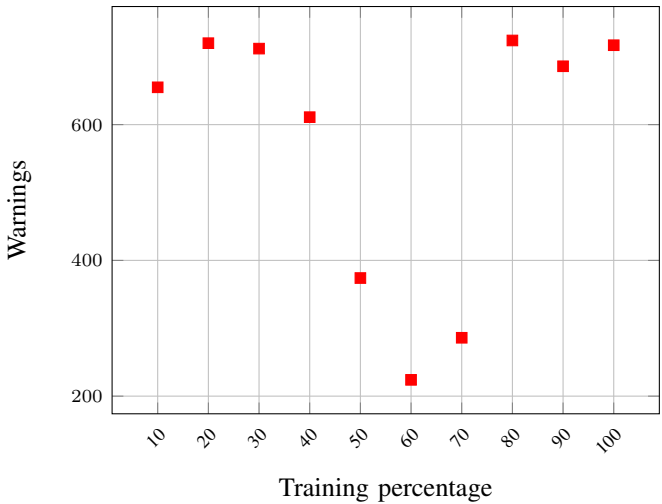
 
\subsection{RQ2}\label{ssec:rq3}
We trained the GTN model with random samples of the dataset at 10\% increments (10\%, 20\%, ...) and ran the evaluation to see how much data is required for good results.
We wanted to present the evaluation of the downstream task instead of presenting standard ML metrics (like F1 score) on their own, which can be potentially misleading: even a classifier that seems to perform well might not be useful for the downstream task.

\Cref{fig_perc} shows the result: a classic diagram of the model improving with training data until 60\%;
performance then worsens after 70\% with overfitting. 
Performance is maximized from training around 50\% of the dataset:
around 16,000 classes.
This estimate gives us a recommendation to calibrate the dataset size for similar tasks.

\subsubsection{Availability of Data for Other Pluggable Type Systems}
\label{sec:rq3-nonnull}

We collected annotations using Sourcegraph's~\cite{b8} src tool to investigate
whether any other pluggable type systems (for Java) besides nullability
have enough publicly-available data to plausibly train them the same way we
trained \NullGTN. We found that none did. The closest were
\<@GuardedBy> (from the Checker Framework's Lock Checker~\cite{ErnstMMS2016})
with 1,752 annotated classes; and integer \<@NonNegative> and \<@Positive> annotations
(from several sources, including the Checker Framework's Index Checker~\cite{KelloggDME2018})
with 1,675 and 1,008 annotated
classes, respectively. These occur an order of magnitude less often than \<@Nullable>,
showing there is insufficient data for us to repeat our experiments with
another pluggable type system until humans annotate more classes:
a ``chicken-and-egg problem'' for type qualifier inference.

\subsubsection{Discussion: ``Chicken-and-Egg Problem'' for Type Qualifier Inference}
\label{sec:chicken-and-egg}

The present work does not try to solve this ``chicken-and-egg problem'' of needing data to enable type qualifier inference (as proposed in this paper),
while human adoption is needed to produce that same data.
Our contribution is to show that simple feature engineering and choice of model is sufficient to do a reasonable job: that is, we have shown the \emph{feasibility} of our approach, if enough data is available.
However, our work does give the community a sense of how much “human adoption” is necessary before training an inference model like \nullgtn is practical,
which we believe is a useful step towards an otherwise intractable problem.
Further, our work shows that once the threshold for ML-based qualifier inference is reached, it can be leveraged to help increase adoption more.

 \section{Limitations and Threats to Validity}
\label{sec:limits-and-threats}

\subsection{Threats to the Validity of Our Experiments}
\label{sec:threats}

One threat to our NullAway experiment (\cref{sec:nullaway-eval})
is that the quality of the human annotations that
we used as ground truth is somewhat questionable: they do not
remove all NullAway warnings on 3 of the projects, which may
distort our results.
Another threat is
the use of proxies for human effort: precision/recall of human-written
annotations and reduction in warning count may both be poor
proxies for effort. For example, even though \NullGTN finds most of
the annotations a human would write, those it fails to find
might be the most difficult or complex to write.
Another threat
comes from the choice of benchmarks: we use programs that
have been, at some point, annotated by a human to typecheck with NullAway, so
that we have ground truth for our recall computation. However,
in a deployment of \NullGTN, we expect that the program would
never have been human-annotated. Programs which have never been
annotated might be trickier for inference, since human annotators might also change code to simplify
the process of annotation (and those changes might stay in the code,
even if the annotations themselves are later removed).
The use of NullAway itself is also
a threat to generalizability: our approach
may not be as effective at inferring annotations that are useful
for some other nullness analysis, such as Meta's NullSafe.

Our approach may not generalize
beyond Java or nullability.
In theory, our approach could be applied to other languages
beyond Java, but evaluating that is out-of-scope.
We expect that the \napast construction procedure in \cref{sec:napast}
and a FastGTN architecture will be applicable
to other pluggable typecheckers that prove the absence
of other problems in Java,
if a dataset of type qualifiers for those checkers were available. However, we cannot test whether
this is effective in practice, since as we showed in \cref{sec:rq3-nonnull} sufficient training data
does not exist for typecheckers that target problems other than nullability:
there is at least an order of magnitude more code annotated for nullability than for any other typechecker
in open-source, and in \cref{ssec:rq3} we showed that \NullGTN required about half
of all the data we collected before it became effective.

Another threat to generalizability is that the data experiment in \cref{ssec:rq3} may
not be accurate for other pluggable type systems. In particular,
nullability type systems, compared to other extant pluggable type systems, are
both relatively simple (only two type qualifiers, in a relatively
simple type hierarchy) and relatively annotation-intensive (because null is pervasive in
most Java programs).
We believe that both of these features probably bias nullability inference to require
\emph{less} data, so our estimate may be overly conservative---type systems with large
type hierarchies, dependent types, or other complex features may require much more
data than our estimates suggest.

\subsection{Generality}
\label{sec:generality}

Our approach works for pluggable type systems with a finite set of types, assuming sufficient training data. The restriction to a finite set of types is an important limitation: we must have fixed the set of type qualifiers to place before training the model, because we’re using a classifier. Many pluggable type systems, such as units-of-measure~\cite{XiangLD2020} or integer signedness (e.g., @NonNegative and @Positive in \cref{sec:rq3-nonnull}), fit this model. However, systems with parameterized qualifiers, like @GuardedBy (\cref{sec:rq3-nonnull}), require a different approach—possibly combining our method (to predict annotations) with a generative model (to generate parameters).

In practice, human-annotated code is an order of magnitude rarer for other checkers than the amount needed to train NullGTN (shown in \cref{sec:rq3-nonnull}). This prevents us from directly evaluating the generality of our approach. We’re investigating approaches to synthesize training data for other pluggable type systems, but that is another, separate problem.

\subsection{Other Limitations of Our Approach}
\label{sec:limits}

All machine-learning-based techniques share one serious limitation:
there are no guarantees that
the annotations that it infers are correct. In our experiments,
\NullGTN demonstrated close alignment with human-written ground
truth annotations---with a recall of 69\%. Though there is also
no guarantee that human-written annotations are correct, we believe
that \NullGTN's ability to hew so closely to developers' annotations
suggests that this problem will be minor in a practical
deployment: whether a human or \NullGTN writes an incorrect annotation,
a sound typechecker will reject it.\looseness=-1

 \section{Related Work}
\label{sec:relatedwork}

The most closely-related work to ours is a recent approach for ``error-driven''
inference of nullability annotations for pluggable typecheckers~\cite{KarimipourPCS2023}.
The key idea of this approach is to treat the typechecker as a ``black box'' and use its
error messages to drive inference: for example, if the checker warns that a return
expression is nullable but that a return type is declared as nonnull, this approach
could infer that an \<@Nullable> annotation should logically be placed on the return
type. In experiments using NullAway, this approach was shown to reduce the number
of warnings emitted by the checker by 69.5\% (vs. our 60\%); they did not measure precision/recall of ground-truth annotations.
However, our techniques are different: they relies on the checker's output, while we rely
on where humans have written nullability
annotations historically (i.e., the training data for our model).
In methodology,
the two approaches are orthogonal; we hope to profitably combine them in future work,
since the differing approaches mean that they probably infer \emph{different} subsets
of nullability annotations.
Another closely related approach bootstraps global inference from
the local inference that is built into most pluggable typechecking frameworks~\cite{KelloggDNAE2023}.
This approach, though not specific to nullness, also uses a fundamentally different approach and might
be profitably combined with ours in future work: it works deductively, unlike our approach.

Some prior works~\cite{XiangLD2020,Li2017} use constraint-based type inference for
pluggable types. These approaches do not support nullness (they are specific
to a units-of-measurement type system), and cannot be readily applied
to other type systems: each is unique.
Our methodology, by contrast, could be applied to any type system,
if training data were available: a model developer would need to repeat our ablation
experiments to drop appropriate nodes from the \napast, but
otherwise could reuse our infrastructure. Prior approaches for explaining type inference
errors~\cite{LernerFGC2007,PavlinovicKW2014,LoncaricCSS2016,ZhangMVPJ2017}
and for gradual type migration~\cite{PhippsCostinAGG2021} suffer the same problems as other constraint-based
approaches (i.e., do not apply easily to new type systems). There is also prior
work that involves the user in type inference~\cite{VakilianPEJ2015}, which is complementary
to ours.
The Daikon invariant detector~\cite{ErnstPGMPTX2007} can insert \<@Nullable>
annotations at program locations that can be null at run time~\cite{daikon-annotatenullable}.
Like any dynamic approach, it is inherently limited by the quality of program's test suite.

There is a lot of work on inference of static types for dynamically-typed
languages such as Python or JavaScript using machine learning~\cite{XuZCPX2016,PradelGLC2020,peng2023statistical,wei2023typet5,peng2023generative,PengGLGLZL2022,mir2022type4py,cui2021pyinfer}.
These approaches are not directly applicable to our problem, since they do not reason
about nullability, but about the base type system, and
their models and encodings are specialized to that problem.

There is a large literature on type inference and type reconstruction algorithms
based on constraint solving~\cite[Chapter 22]{pierce:2002:types-and-pls}. These techniques
build and then solve a system of constraints; the
specific constraints are type-system-dependent, which makes this approach unsuitable
for pluggable types: determining the constraints to solve would require translating
the typechecker's logic into the logic of the constraint solver, which effectively would
require re-implementing the typechecker.
Recent work~\cite{MadsenP2021} has proposed an extension to a classic constraint-based
type inference approach~\cite{DamasM1982} that enables
reasoning about nullability. However, extending this to the mainstream object-oriented
languages (and their associated nullability checkers) is an open-problem.

\section{Data Availability}
\label{sec:data}
The data and code from our experiments are open-source and available~\cite{our-data}.

\bibliographystyle{ACM-Reference-Format}
\bibliography{missing-annos,plume-bib/ernst,plume-bib/types,plume-bib/sideeff,plume-bib/soft-eng,plume-bib/crossrefs}

\end{document}